\documentclass[aip,graphicx]{revtex4-1}
\usepackage{graphicx}
\usepackage{subcaption}
\usepackage{float}
\usepackage{todonotes}

\draft 

\begin{document}


\title{Efficient heating of single-molecule junctions for thermoelectric studies at cryogenic temperatures} 



\author{Pascal Gehring}
\email{p.gehring@tudelft.nl}
\affiliation{\small \textit Kavli Institute of Nanoscience, Delft University of Technology, Lorentzweg 1, 2628 CJ Delft, The Netherlands}

\author{Martijn van der Star}
\affiliation{\small \textit Kavli Institute of Nanoscience, Delft University of Technology, Lorentzweg 1, 2628 CJ Delft, The Netherlands}

\author{Charalambos Evangeli}
\affiliation{\small \textit Department of Materials, University of Oxford, Parks Road, OX1 3PH, Oxford, United Kingdom}
\affiliation{\small \textit Department of Physics, Lancaster University, Bailrigg, LA1 4YB, Lancaster, United Kingdom}

\author{Jennifer J. Le Roy}
\affiliation{\small \textit Department of Materials, University of Oxford, Parks Road, OX1 3PH, Oxford, United Kingdom}

\author{Lapo Bogani}
\affiliation{\small \textit Department of Materials, University of Oxford, Parks Road, OX1 3PH, Oxford, United Kingdom}

\author{Oleg V. Kolosov}
\affiliation{\small \textit Department of Physics, Lancaster University, Bailrigg, LA1 4YB, Lancaster, United Kingdom}

\author{Herre S. J. van der Zant}
\affiliation{\small \textit Kavli Institute of Nanoscience, Delft University of Technology, Lorentzweg 1, 2628 CJ Delft, The Netherlands}



\date{\today}

\begin{abstract}
The energy dependent thermoelectric response of a single molecule contains valuable information about its transmission function and its excited states. However, measuring it requires devices that can efficiently heat up one side of the molecule while being able to tune its electrochemical potential over a wide energy range. Furthermore, to increase junction stability devices need to operate at cryogenic temperatures. In this work we report on a new device architecture to study the thermoelectric properties and the conductance of single molecules simultaneously over a wide energy range. We employ a sample heater in direct contact with the metallic electrodes contacting the single molecule which allows us to apply temperature biases up to $\Delta T = 60$~K with minimal heating of the molecular junction. This makes these devices compatible with base temperatures $T_\mathrm{bath} <2$~K and enables studies in the linear ($\Delta T \ll T_\mathrm{molecule}$) and non-linear ($\Delta T \gg T_\mathrm{molecule}$) thermoelectric transport regimes.

\end{abstract}


\maketitle

Theory predicts that electrical and thermoelectric properties of single molecules can be tailored by chemical design. For example, adding pendant groups to a conjugated molecule backbone can introduce sharp features in its energy dependent transmission probability, because of quantum interference effects,\cite{Valkenier2014} and such sharp features should generate exceptionally high thermoelectric efficiencies\cite{Finch2009}. Furthermore, single molecules can host a rich variety of physical effects:\cite{Gehring2019} strong electron-phonon interactions\cite{Koch2006}, strong correlations and Kondo effects\cite{Liang2002}, or exotic blockade phenomena.\cite{Bruijckere2019} All these are predicted to strongly influence the thermoelectric properties\cite{Koch2004,Sowa2019,Wang2010,Costi2010}, but these predictions remain untested, because of a lack of appropriate experimental platforms. 

In order to perform detailed thermoelectric characterisations of single molecules, the device architecture needs to fulfill the following conditions: the device needs to be compatible with methods to contact single molecules; a gate electrode is necessary for a full characterisation of the thermoelectric properties of the single-molecule junction; because of the thermal instabilities in molecular junctions, the devices need to be compatible with cryogenic temperatures; and for the same reason the temperature difference between the hot and the cold side $\Delta T = T_\mathrm{hot} - T_\mathrm{cold}$ in the molecular junction must not heat excessively the molecule itself. So far, only a few device architectures exist that fulfill some of the aforementioned conditions, based on graphene\cite{GehringNL2017} or Au electrodes\cite{Kim2014}. These devices suffer, however, from low heating efficiencies ($50 - 150$~mK~mW$^{-1}$) and, in devices with a side heater, the temperature profile along the channel is approximately linear\cite{Moon2013} so that high heater powers are necessary to apply $\Delta T$ across short junctions. For the case of graphene junctions, a side heater produces strong heating of the cold side of the junction, characterized by $(T_{\mathrm{cold}}-T_{\mathrm{bath}})/\Delta T \approx 5$, where $T_{\mathrm{bath}}$ is the temperature of the cryostat. This makes these devices not compatible with measurements at low cryogenic temperatures.

Here, we develop a novel device architecture for simultaneously studying the electric and thermoelectric properties of single molecules as a function of the gate voltage $V_g$. Fabrication is based on electromigration and self-breaking of Au, leading to several key advantages: Au enables access to different tunnel coupling strengths (e.g. by using thiol bonds, and spacer linkers)\cite{Su2016}; self-breaking\cite{ONeill2007} of the Au bridges can prevent the formation of spurious quantum dots, which is sometimes a problem for carbon-based leads\cite{GehringNL2016}; the close proximity of the sample heater to the leads enables efficient heat transfer, while reducing heating of the single molecule at temperature $T_\mathrm{molecule} = (T_\mathrm{hot} + T_\mathrm{cold})/2$ (thus ensuring device stability) and enabling experiments at $T_\mathrm{bath} < 2$~K. The improved heating efficiency also provides access to a wide $\Delta T$ range (mK to few tens of K), opening the way to the study of the thermoelectric properties of single-molecule junctions in the linear ($\Delta T \ll T_\mathrm{molecule}$) and non-linear ($\Delta T \gg T_\mathrm{molecule}$) thermal bias regimes. Moreover, this novel method allows the simultaneous measurement of the gate-dependent conductance $G(V_\mathrm{g})$ and thermoelectric current $I_\mathrm{th}(V_\mathrm{g})$. This eliminates the problem that small drifts of the signals (because of hysteresis effects of the gates or activation of charge traps in the gate oxide) can hinder a direct comparison of data sets when the two quantities are measured subsequently, as in previous devices.\\

\begin{figure}[h!]
    \centering
    \includegraphics[width = 1\textwidth]{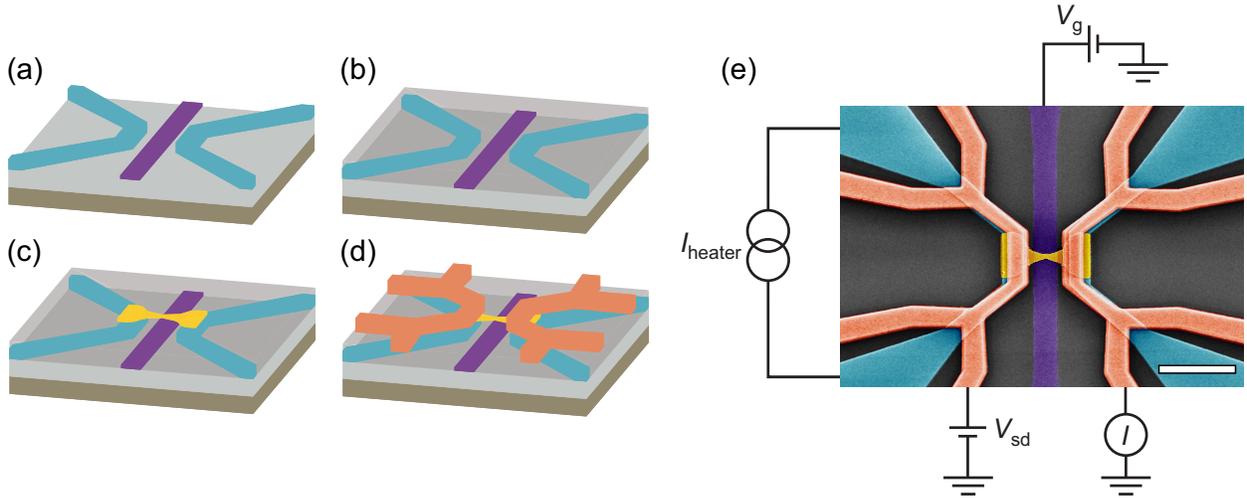}
    \caption{(a) - (d) Overview of the fabrication process. (a) Fabrication of the local back gate electrode (purple) and sample heaters (blue). (b) Deposition of a thin Al$_2$O$_3$ insulating layer on top of the whole device. (c) Deposition of a Au bridge (yellow) which is (d) contacted by two four-terminal thermometers (orange). (e) False-color scanning electron micrograph of the single-molecule transistor architecture, consisting of a thin Au bridge (yellow) on top of a gate electrode (purple) connected by two four-terminal thermometers (orange) which are on top of the sample heaters (blue). The schematic circuit diagram indicates the terminals used for $G$ and $I_{th}$ measurements: a source-drain voltage $V_\mathrm{sd}$ can be applied to the drain while a current to ground $I$ is measured at the source. $V_\mathrm{g}$ is applied via the back gate with respect to ground. Heater currents $I_\mathrm{heater}$ are applied to the sample heater. Scale bar: 2~$\mu$m.}
    \label{fig:fig1}
\end{figure}

The junctions are fabricated following the scheme depicted in Figures \ref{fig:fig1}(a) - (d). A Pd sample heater (3~nm Ti/27~nm Pd) and Pd gate electrode (1~nm Ti/6~nm Pd) were patterned on a Si wafer with 817~nm SiO$_2$ using standard electron beam lithography and electron beam evaporation (Figure \ref{fig:fig1}a). A thin gate electrode is used to reduce thermal transport between drain and source lead. Pd is used because it is known to form uniform thin layers with low surface roughness.\cite{NAZARPOUR20105715} In a second step a 10~nm Al$_2$O$_3$ insulating layer is globally applied by atomic layer deposition (Figure \ref{fig:fig1}b). This layer serves as a gate dielectric and as an insulation layer to electrically insulate the sample heater from the drain and source leads.\cite{Gluschke2014} Thereafter, a 12~nm thick bow-tie shaped Au bridge (narrowest part $<60$~nm) is evaporated (Figure \ref{fig:fig1}c) and electrically contacted by two four-terminal thermometers (5~nm Ti/65~nm Au, Figure \ref{fig:fig1}d). The effective temperature drop on a molecule trapped between the two Au contacts depends on the thermal resistances of the Au bridge. Therefore a short channel length should be used to reduce its thermal resistance which ensures thermalisation with the heated Au contact. On the other hand, very short channels promote direct heating of the 'cold' contact by the sample heater. In this study we chose a short channel length of $1$~$\mu$m. Figure \ref{fig:fig1}e shows a false-color scanning electron microscopy image of a final device. To use these devices for studying the thermoelectric properties of single molecules we open a nm sized gap in the Au bridge by electromigration\cite{Park1999} followed by self-breaking\cite{ONeill2007} to avoid the formation of Au clusters inside the junction.

In the following we describe the methods for estimating $\Delta T$ created by the sample heater after electromigration. We employed two calibration techniques: Scanning thermal microscope (SThM) mapping in high vacuum and resistance thermometer method using the drain and source contacts as thermometers. For the former, we used a home-built high vacuum SThM\cite{Pumarol2012} with commercially available (Anasys Instruments, AN-300) doped silicon probes which are geometrically similar to standard micromachined AFM probes. The probe temperature $T_\mathrm{probe}$ can be controlled with an integrated heater at the end of the cantilever, which also acts as a temperature sensor when the tip is in contact with the sample. The electrical response of the probe heater as a function of excess mean probe temperature ($\Delta T_\mathrm{probe} = T_\mathrm{probe} - T_\mathrm{bath}$) was calibrated on a heated stage inside the high vacuum chamber, following a procedure described elsewhere\cite{Tovee2012}.

\begin{figure}[h!]
    \centering
    \includegraphics[width = 0.5\textwidth]{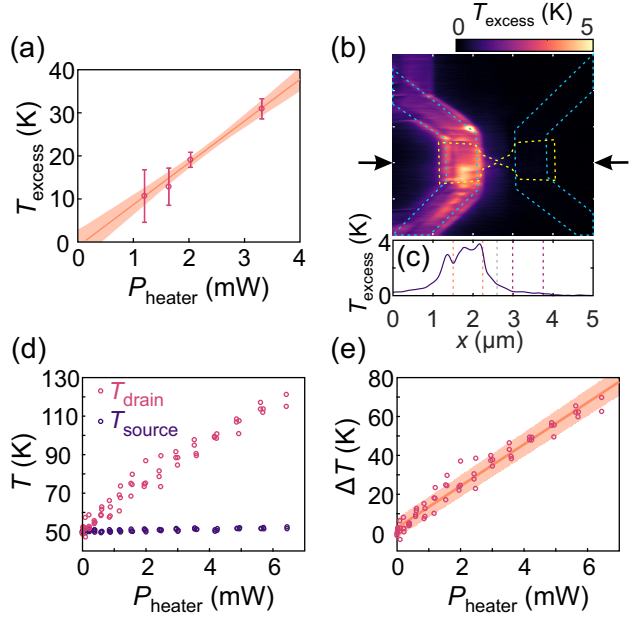}
    \caption{(a) Results of the SThM null-point method. Excess temperature $T_\mathrm{excess}$ of the drain (hot) lead as a function of heater power. The error of the linear fit is indicated by the red shaded area. (b) Temperature map of the device recorded using non-equilibrium thermometry method at $P_{\mathrm{heater}}=0.38$~mW. The dotted lines indicate the position of the drain and source leads, and the gold bridge, respectively. A line cut along the device (indicated by arrows) is shown in (c). (d),(e) Results of the calibration using the resistance thermometer method. (d) Temperature of the drain and source lead as a function of heater power. (e) Temperature drop $\Delta T = T_\mathrm{drain} - T_\mathrm{source}$ across the junction as a function of heater power. The red shaded area indicates the error of the linear fit.}
    \label{fig:fig2}
\end{figure}

Two different quantitative SThM methods were employed to estimate $\Delta T$: the null-point method\cite{Hwang2014} and a non-equilibrium thermometry method\cite{Menges2016,Harzheim2018}. In the null-point method the probe is brought into contact with the sample for different $T_\mathrm{probe}$ while the SThM response is recorded. A jump in the SThM response signal is typically observed at the tip-sample mechanical contact when the probe apex and sample are at different temperatures (examples in Figure S1, Supporting Information). The jump is positive/negative when the temperature of the probe apex, $T_\mathrm{apex}$, is larger/smaller than that of the sample, $T_\mathrm{sample}$, and zero when they are the same. $T_\mathrm{apex}$ in contact with the sample has been found\cite{Tovee2012} to be 88$\%$ of $T_\mathrm{probe}$. Using this procedure, we measured the $T_\mathrm{excess} = T_\mathrm{sample}- T_\mathrm{bath}$ of the drain (hot) lead for 4 different powers applied to the sample heater which is plotted in Figure \ref{fig:fig2} (a). Linear regression yields a conversion factor of $9.8 \pm 1.2$~K~mW$^{-1}$, with an error originating mainly from the temperature calibration of the probe and the estimation of the jump of the SThM signal (see Supporting Information), especially for low $T_\mathrm{probe}$ where the SThM signal noise is comparable to the signal jump. 

The second SThM method relies on non-equilibrium thermometry where an AC bias voltage is applied to the sample heater and the resulting variations of $T_\mathrm{excess}$ are detected by the SThM tip. The $T_\mathrm{excess}$ map is extracted through the relation $T _\mathrm{excess}=\Delta T _\mathrm{probe} \frac{\Delta V_\mathrm{AC}}{\Delta V_\mathrm{AC}-\Delta V_\mathrm{DC}}$, where $\Delta V_\mathrm{AC}$ is the AC SThM response detected at the second harmonic and $\Delta V_\mathrm{DC}$ the DC SThM signal due to heat flux from the sample to the tip. Modulation of the sample heater with high frequencies can lead to damping of the SThM signal since thermal equilibrium can only be reached within a time scale $\tau_\mathrm{th}$ given by the total thermal capacitance and all thermal resistances of our device. For the temperature mapping, a modulation frequency of 7~Hz is used, due to limitations in the lowest possible scanning speed, which is slightly bigger than $1/\tau_\mathrm{th}$ and which results in a reduction of SThM signal by about 10$\%$. We account for this damping by rescaling of the $T_\mathrm{excess}$ maps using Figure \ref{fig:fig4} (c). The resulting map for a device with $P_\mathrm{heat}=0.38$~mW applied to the sample heater is shown in Figure \ref{fig:fig2} (b).

From this $T_\mathrm{excess}$ map and a line cut through this map in Figure \ref{fig:fig2} (c) we observe that for a heating of the hot (left) contact by about 3~K the cold (right) contact only heats up by about 0.14~K, which yields a very low $(T_{\mathrm{cold}}-T_{\mathrm{bath}})/\Delta T \approx 0.05$. This low heating of the cold side allows us to estimate $\Delta T$ from the excess temperature of the drain lead in Figure \ref{fig:fig2} (a) using $\Delta T \approx T_\mathrm{excess}$. It is worth to mention that the temperature of the gold bridge differs noticeably from that of the drain and source contacts. This has been observed in previous studies\cite{Kim2014} and would result in an overestimation of $\Delta T$ across the molecule in the centre of the junction. However, SThM only accesses the phonon (lattice) temperature $T_\mathrm{ph}$, and the electron temperature $T_\mathrm{e}$ (which drives thermoelectric effects) can be much higher when using efficient sample heater in direct contact with leads\cite{Gluschke2014}. Since we cannot access the real drop in $T_\mathrm{e}$ on the single-molecule junction, in the remainder of this paper we use the $\Delta T$ between the drain and source lead for calculations, which leads to an underestimation of the thermoelectric coefficients and efficiencies.

The second technique used to estimate $\Delta T$ is the resistance thermometer method.\cite{Kim2001,SMALLssc2003,Smallprl2003} To this end, we use the four contacts connecting the drain and source lead to first measure their 4-terminal resistance as a function of $T_\mathrm{bath}$ in a cryostat. 
Thereafter the sample temperature is held constant (here $T_\mathrm{bath}=50$~K) and the 4-terminal resistance of the drain and source leads are measured as a function of dissipated heater power. Combining of both measurement results allows estimating $T_\mathrm{drain}$ and $T_\mathrm{source}$ as a function of heater power $P_\mathrm{heater}$ (see Figure \ref{fig:fig2} (d)). It can be seen that the (hot) drain lead in direct contact with the sample heater heats up by tens of Kelvin when increasing the heater power while the (cold) source lead stays almost at $T_\mathrm{bath}$. Using this data we estimate $\Delta T$ as a function of $P_\mathrm{heater}$ (Figure \ref{fig:fig2} (e)). We find that $\Delta T$ increases linearly with $P_\mathrm{heater}$, which allows to accurately apply small $\Delta T$ biases. Extracting the slope of 10.7~$\pm$~0.8~K/mW, we find a heating efficiency of $\Delta T/(P_{\mathrm{heat}}L) = 10.7 \pm 0.8$~K~mW$^{-1}$~$\mu$m$^{-1}$ at 50~K. This value is close to the value found using the SThM methods above. 

The efficiency found in our devices is orders of magnitude higher than that found in devices with side heaters\cite{GehringNL2017} and it is comparable to similar devices designed to study thermoelectric properties of nanowires which use sample heater patterned on top of the leads\cite{Gluschke2014}. Such a high heating efficiency allows to drive systems into the non-linear regime where $\Delta T$ becomes comparable to, or even exceeds $T_\mathrm{bath}$. This is demonstrated in Figure \ref{fig:fig2} (e) (which was recorded at $T_\mathrm{bath} = 50$~K) for $P_\mathrm{heater} > 5$~mW, where $\Delta T > 50$~K. Moreover, from the data in Figure \ref{fig:fig2} (d) we find a low $(T_{\mathrm{cold}}-T_{\mathrm{bath}})/\Delta T < 0.026$, which indicates minimal heating of the cold reservoir and the molecule. This value, which is significantly lower than previously-reported values\cite{GehringNL2017,Gluschke2014}), ensures stability of the molecular junction and enables experiments at $T_\mathrm{bath}<2$~K.\\

In the following we test the device architecture to measure the thermocurrent of a single [Gd(tpy-SH)$_2$(NCS)$_3$] molecule. by immersing the sample in a 0.5~mM molecule solution in dichlormethan after electromigration and self breaking. We observe molecular junction formation indicated by occurrence of gate dependent transport features at $T_\mathrm{bath} = 1.8$~K in 7 out of 47 junctions. This junction formation yield of $\approx 15 \%$ is similar to values that we typically observe for electromigrated Au electrodes.\cite{Burzuri2015} In this paper we focus on demonstrating the suitability of our junctions for thermoelectric characterisation of single molecules and present the data for one selected device.

\begin{figure}[h!]
\centering
    \includegraphics[width=1\textwidth]{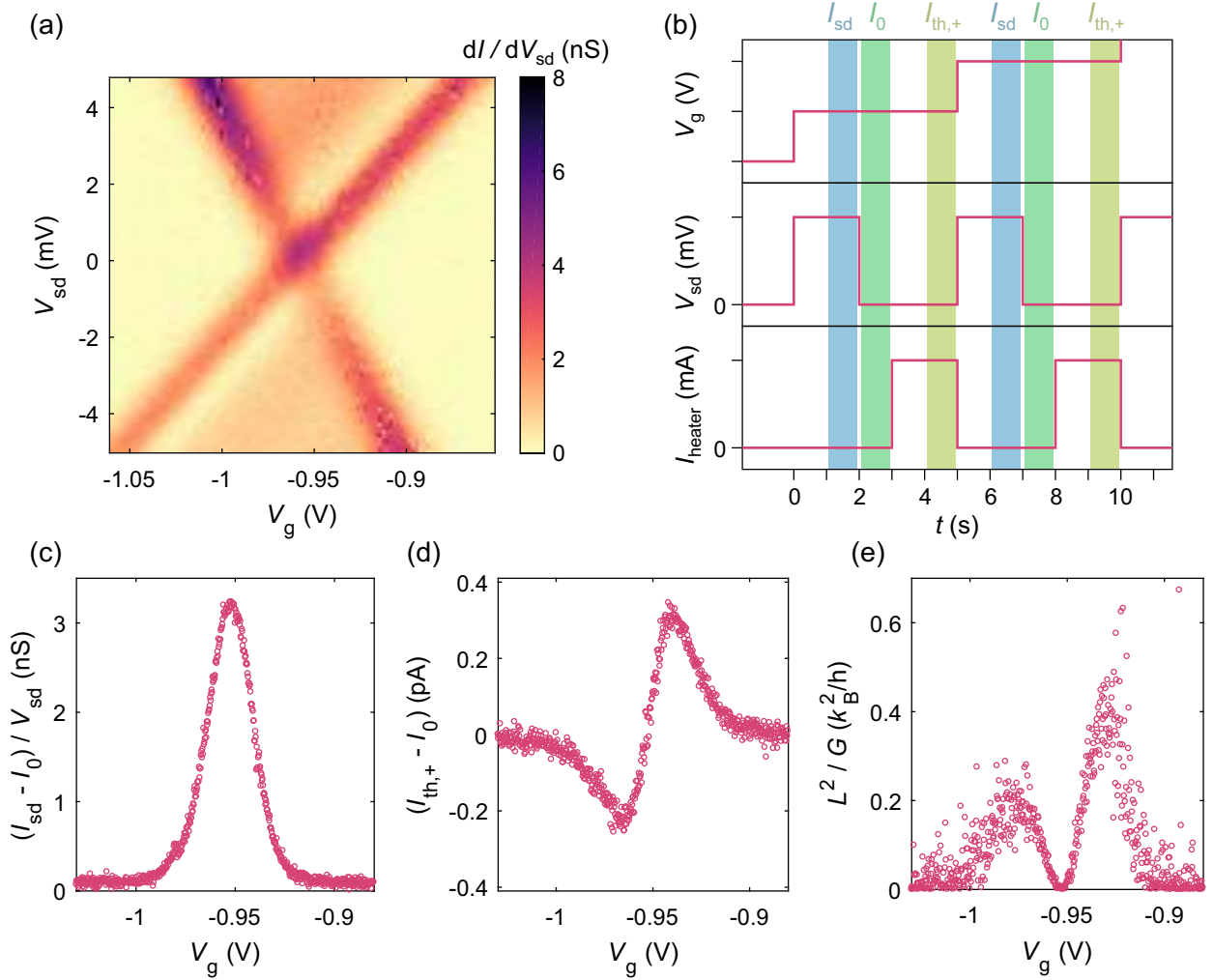}

    \caption{(a) Differential conductance $\mathrm{d}I/\mathrm{d}V_\mathrm{sd}$ as a function of applied gate $V_\mathrm{g}$ and bias $V_\mathrm{sd}$ voltages. (b) Measurement scheme used to measure $I_\mathrm{th}(V_\mathrm{g})$ and $G(V_\mathrm{g})$ as a function of gate voltage. The shaded regions indicate the time windows in which current measurements are performed. (c) Conductance, (d) $I_\mathrm{th}$ and (e) power factor as a function of $V_\mathrm{g}$.}
    \label{fig:fig3}
\end{figure}

Figure \ref{fig:fig3} (a) shows the differential conductance $\mathrm{d}I/\mathrm{d}V_\mathrm{sd}$ of a molecular junction as a function of bias voltage $V_\mathrm{sd}$ and $V_\mathrm{g}$. Two regions with low $\mathrm{d}I/\mathrm{d}V_\mathrm{sd}$ (yellow) are separated by two crossing lines of high $\mathrm{d}I/\mathrm{d}V_\mathrm{sd}$. These lines are attributed to the borders of so-called Coulomb diamonds. The current inside the two adjacent diamonds is suppressed due to Coulomb blockade, whereas sequential electron tunneling occurs inside the hour-glass shaped region.\cite{Gehring2019}

$I_\mathrm{th}$ and $G$ were then measured simultaneously in the device configuration shown in Figure \ref{fig:fig1} (e) following the measurement protocol depicted in Figure \ref{fig:fig3} (b). $V_\mathrm{g}$ is first ramped to the desired value and a small $V_\mathrm{sd} = 0.5$~mV is applied. After a short settling time $I_\mathrm{sd}$ is measured, $V_\mathrm{sd}$ is set to zero and a offset current $I_0$ may be measured, which can originate from gate leakage currents or offsets in the current pre-amplifier. Subsequently, a heater current $I_\mathrm{heater} = 0.1$~mA ($P = 2.6 \mu$W) is applied to the sample heater, followed by a settling time and a measurement of the raw thermocurrent, $I_\mathrm{th,+}$. These measurement steps are repeated for each gate voltage value. Using the three measured current values the conductance $G=(I_\mathrm{sd}-I_0)/V_\mathrm{sd}$ and the thermocurrent $I_\mathrm{th}=I_\mathrm{th,+}-I_0$ are calculated. The power factor $S^2G = (V_\mathrm{th}/\Delta T)^2 G = (I_\mathrm{th}/\Delta T)^2/G$, which is a measure for the amount of energy that can be generated from a certain $\Delta T$, is thus determined directly. 

Figure \ref{fig:fig3} (c) and (d) show the results of this measurement on the molecular junction. The conductance $I_\mathrm{sd}/V_\mathrm{sd}$ peaks at around $V_\mathrm{g} = -0.96$~V. This indicates the energetic position of the charge degeneracy point where the transition from the $N$ to the $N+1$ charge state of the molecule occurs (corresponds to closing point of the Coulomb diamonds in Figure \ref{fig:fig3} (a)). Furthermore, we extract the gate coupling factor $\alpha = C_\mathrm{g}/(C_\mathrm{s}+C_\mathrm{d}+C_\mathrm{g}) = 33$~meV/V, from the slopes of the Coulomb diamond following Ref. \citenum{Hanson2007}. This gate coupling factor, which is a factor 4-5 higher than the typical values found for devices using SiO$_2$ back gates\cite{Osorio2008, GehringACS2017}, enables efficient tuning of the single-molecule junction and allows thermoelectric studies over a wide energy range, of about $\pm 400$~meV, as estimated using the typical break down voltages of 12-14~V found in our devices.

Figure \ref{fig:fig3} (d) shows $I_\mathrm{th} = I_\mathrm{th,+}-I_0$ as a function of $V_\mathrm{g}$, displaying a resulting curve that is S-shaped and changes sign at the charge degeneracy point. This sign change indicates that the transition from electron- to hole-like thermocurrents occurs when crossing the charge degeneracy point, in agreement with theoretical predictions and previous experiments.\cite{RinconGarcia2016,Gehring2019} By tuning the system far away from resonance, $I_\mathrm{th}$ vanishes. Combining the data in Figure \ref{fig:fig3} (c) and (d) and using $\Delta T \approx 30$~mK obtained from our calibration allows calculating the gate-dependent power factor $S^2G = L^2/G$, where $S = -V_\mathrm{th}/\Delta T$ is the Seebeck coefficient, $V_\mathrm{th}$ is the thermovoltage and $L = -I_\mathrm{th}/\Delta T$ is the thermal response coefficient. The result of this calculation is shown in Figure \ref{fig:fig3} (e). The power factor can be tuned from zero to about $0.4 k_\mathrm{B}^2 /h$, which is close to the theoretical limit of $(1/2.2) k_\mathrm{B}^2 /h$ predicted for a single quantum level.\cite{GehringNL2017}

In the remainder of this paper we test if the device platform developed in this study is suitable for AC thermoelectric measurements.\cite{SMALLssc2003} For this purpose an AC current at frequency $f$ is applied to the sample heater and $I_\mathrm{th}$ is measured at the second harmonic, $2f$. As can be shown\cite{Zuev2011} the maximum signal in the second harmonic is at a phase of $90^{\circ}$ with respect to the excitation. Furthermore, the raw data needs to be multiplied by a factor of $2\sqrt{2}$ to convert it from rms to peak-to-peak and to correct the shift in reference when locking to the second harmonic.\cite{GehringNL2017} Figure \ref{fig:fig4} (a) shows the AC thermocurrent as a function of gate voltage measured with $f=3$~Hz for the same device discussed above. 
\begin{figure}[h!]
\centering
    \includegraphics[width=0.5\textwidth]{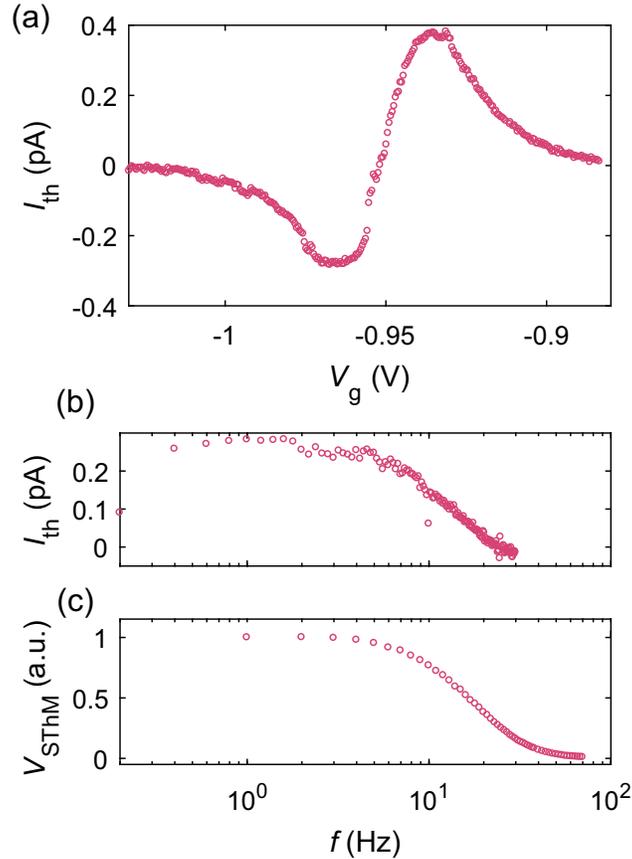}

    \caption{(a) AC thermocurrent ($f=3$~Hz, $I_\mathrm{heater} = 0.1$~mA, $P_\mathrm{heater} = 26$~$\mu$W) as a function of gate voltage. (b) Thermocurrent at $V_\mathrm{g}=-0.965$~V as a function of modulation frequency of the sample heater. (c) SThM signal on the drain (hot) contact as a function of modulation frequency of the sample heater.}
    \label{fig:fig4}
\end{figure}
Line shape and amplitude of the AC measurement match the results of the DC measurement in Figure \ref{fig:fig3} (d) well. This changes if higher frequencies are used: in Figure \ref{fig:fig4} (b) the AC thermocurrent measured at fixed gate voltage ($V_\mathrm{g}=-0.965$~V) as a function of modulation frequency of the sample heater is shown. Above a frequency of about 3~Hz the signal amplitude drops from its DC value to zero when reaching frequencies of about 30~Hz. This can be explained by the thermal equilibrium time of the system as discussed above. To illustrate this the SThM signal measured on the hot contact as a function of sample heater excitation frequency is shown in Figure \ref{fig:fig4} (c). A similar trend as for the thermocurrent signal can be observed where a deviation from the DC signal strength occurs at $f > 3$~Hz.

In summary, we developed a new device architecture and the first robust measurement protocol that allows measuring the thermoelectric properties of single molecules at cryogenic temperatures, over a wide energy range. The close proximity of the sample heater to the electrical contacts yields a high heating efficiency and low global heating of the molecular junction itself. This ensures device stability and allows to accurately study thermoelectric effects over wide $\Delta T$ ranges. Furthermore, we demonstrate that the gate dependent thermocurrent and conductance can be measured in parallel and that the devices are suitable for AC measurements, if the excitation frequency is chosen to be smaller than the thermal response time of the system. The devices presented in this study could thus be readily used to study the thermoelectric properties of single molecules in the non-linear regime\cite{Sanchez2013} or to investigate the thermoelectric response of single-molecule magnets\cite{Wang2010} or high-spin molecules in the Kondo regime\cite{Costi2010}. What is more, the Gd-based molecules used in this study are promising candidates for observing single-molecule magneto-cooling effects\cite{Karotsis2009} which are now within experimental reach. 

\begin{acknowledgments}
This work was supported by the EC H2020 FET Open project 767187 “QuIET” and ERC (StG-OptoQMol-338258 and CoG-MMGNRs-773048). P.G. and J.R. acknowledge Marie Skłodowska-Curie Individual Fellowships (Grant TherSpinMol-748642 and SpinReMag-707252) from the European Union’s Horizon 2020 research and innovation programme.
\end{acknowledgments}

\bibliography{FabAPLMartijn}

\end{document}